\documentclass[12pt]{article}

\pdfoutput=1

\usepackage{jcappub}
\usepackage{bm}
\usepackage{amssymb,amsmath,amsthm}
\usepackage{graphics}

\def\widebar{\overline}

\begin{document}
\title{Attracted to de Sitter II:\\ cosmology of the shift-symmetric Horndeski models}
\author[a,b]{Prado Mart\'{\i}n-Moruno}
\author[a]{{\rm and} Nelson J.~Nunes}
\affiliation[a]{Instituto de Astrof\'isica e Ci\^encias do Espa\c{c}o, Universidade de Lisboa, 
Faculdade de Ci\^encias, Campo Grande, PT1749-016 Lisboa, Portugal;}
\affiliation[b]{Departamento de F\'{\i}sica Te\'orica I, Universidad Complutense de Madrid, E-28040 Madrid,
Spain.}
\emailAdd{pradomm@ucm.es}
\emailAdd{njnunes@fc.ul.pt}


\abstract{
Horndeski models with a de Sitter critical point for any kind of material content may
provide a mechanism to alleviate  the cosmological constant problem. 
Moreover, they could allow us to understand the current accelerated expansion of the universe as the result of the dynamical approach to the critical point when it is an attractor. 
We show that this critical point is indeed an attractor for the shift-symmetric subfamily of models with 
these characteristics.
We study the cosmological scenario that results when considering radiation and matter content, 
and conclude that their background dynamics is compatible with the latest observational data.

\noindent\today
}

\maketitle


\section{Introduction}

The cosmological constant problem \cite{Weinberg:1988cp,Carroll:2000fy} is one of the main open problems in theoretical physics.
This problem arises not only due to the fine tuning necessary to (partially or completely) cancel the value of the vacuum energy, but also due to the need of repeating such a fine tuning
whenever a phase transition occurs changing the value of this vacuum energy by a finite amount or due to radiative corrections (see, for example, reference \cite{Kaloper:2014dqa}). 
Moreover, the cosmological constant problem is present in any theory of gravity in which vacuum energy gravitates.
One can think of a number of  possible ways of cancelling the contribution of the vacuum energy in the dynamics of the universe,  however,  as Weinberg realised, this is a non-trivial task \cite{Weinberg:1988cp}.

One of the possible stances towards this problem consists of admitting the existence of some unknown symmetry that fixes the value of the vacuum energy permanently to zero, and
consider alternative theories of gravity to describe the current acceleration of the Universe.
The simplest extension of general relativity consist in considering a four-dimensional metric theory of gravity with a scalar field  non-minimally coupled to the metric.
The form of the Lagrangian must  be constrained in order to avoid the Ostrogradski instability, thus, one can consider a  Lagrangian that only depends on first derivatives of the field. 
Galileon fields, however, are an example of a field Lagrangian containing second order derivatives 
which lead to a second order field equation \cite{Nicolis:2008in}.
The covariantization of this Lagrangian in curved spacetime  with a field which is shift-symmetric was, therefore, a promising model \cite{Deffayet:2009wt}. 
However, this covariant galileons are strongly constrained
by gravitational probes \cite{Barreira:2013eea,Barreira:2013jma,Barreira:2013xea}.

Generalized galileons \cite{Deffayet:2011gz} have evolved beyond the covariantization of reference \cite{Deffayet:2009wt} and comprise models with the most general Lagrangian containing second order derivatives of the field
and leading to second order field equations, without assuming any symmetry for the field. 
The resulting Lagrangian has been proven \cite{Kobayashi}
to be equivalent to the already known (although also slightly forgotten) 
Horndeski theory \cite{Horndeski:1974wa}. 
In the context of this theory one could look for an adjustment mechanism able to alleviate the cosmological constant problem.
The fab four models \cite{Charmousis:2011bf,Charmousis:2011ea} avoided Weinberg's no go theorem by relaxing one of its assumptions. The resulting models have a scalar field which is able to self-tune,
that is, the field screens the contribution of any value of the vacuum energy leading to Minkowski. 
Considering that such a self-tuning has to be dynamical to allow non-trivial cosmological dynamics, one is led to the conclusion that the fab four cosmologies have a Minkowski critical point for any value
of the vacuum energy and any kind of matter \cite{Copeland:2012qf}.

In reference \cite{Martin-Moruno:2015bda}, the concept of self-tuning was extended to consider that the field screens the vacuum energy leading to a de Sitter instead of a Minkowski vacuum. 
Thus, in this scenario, the current accelerated expansion of the universe results from the dynamical 
approach to a stable critical point.
It has been emphasized that the value of $\Lambda$ characterizing the critical point is completely 
independent of the vacuum energy and, therefore, its value is unaffected by phase transitions.
However, in order to alleviate the cosmological constant problem 
it would be necessary that not only the critical point is an attractor but also that a long enough 
radiation and matter phases can be described before 
the attractor is approached, even if the vacuum energy takes a huge value.

As found in reference \cite{Martin-Moruno:2015bda} there are two families of models able to self-tune to a spatially flat de Sitter vacuum. The first family has a minisuperspace Lagrangian density that
is proportional to the derivative of the field. 
The cosmological consequences of these linear models have been studied in reference \cite{Martin-Moruno:2015lha}. 
Although promising results have been presented, the considered models were not completely able to describe the background cosmology of our Universe,
nor results concerning the stability of the critical point can be obtained in the general linear case.

The second family of models has a minisuperspace Lagrangian with a non-linear dependence on the temporal 
derivative of the field which vanishes when evaluated at the Hubble parameter characterizing the critical 
point. These models contain three arbitrary functions of the field and its derivatives. 
Focusing our attention on shift symmetric functions can simplify the general treatment of this 
family of models.
Moreover, it has been proven that the galileon interactions are nonrenormalizable
and that the additional terms generated by quantum corrections contributing to the Lagrangian
are strongly suppressed at energies below the cut-off \cite{Nicolis:2004qq, Nicolis:2008in,deRham:2012ew}.
Therefore, we expect that the screening mechanism, 
which is a result of the particular form of the Lagrangian \cite{Martin-Moruno:2015bda},
would not be spoiled by field and/or matter quantum corrections for the shift-symmetric models
(which are curved spacetime generalizations of the galileon).
It must be noted that by assuming a shift-symmetric field, 
one is not automatically in the covariant galileon case ruled out by observations \cite{Barreira:2013eea,Barreira:2013jma,Barreira:2013xea}
and, therefore, these models are interesting and worthy of being explored. 
In this article we study the cosmology of the shift-symmetric non-linear Horndeski models with a de Sitter critical point regardless of the material content.

This article can be outlined as follows:
In section \ref{model} we summarize some characteristics of the Horndeski cosmological models able to screen any value of the cosmological
constant to a given de Sitter space, focusing attention on the case of a shift-symmetric field.
In section \ref{dynamical} we study the dynamical system for these models showing that the de Sitter critical point is always an attractor
if the material content of the universe satisfies the null energy condition.
In section \ref{hlargo} we consider some simple cases which allow us to gain some intuition regarding the background cosmology of these models.
This study allows us to consider a particular satisfactory model which we present in section \ref{models}.
In section \ref{discussion} we summarize the main conclusions of the article.
As the explicit form of the general Lagrangian of the satisfactory models can be of great interest in order to consider further physical scenarios, we include in appendix \ref{Hfunctions} and \ref{Dfunctions} the functions which appear in this Lagrangian according to the
notation of reference \cite{Horndeski:1974wa} and \cite{Deffayet:2011gz}, respectively.

\section{The non-linear models}\label{model}
The minisuperspace Lagrangian of the Horndeski models with a spatially flat de Sitter critical point for any value of the vacuum energy can be 
expressed as \cite{Martin-Moruno:2015bda}
\begin{equation}
 L=L_{\rm EH}+L_{\rm linear}+L_{\rm nl}+L_{\rm m}.
\end{equation}
In this Lagrangian we have explicitly written the Einstein--Hilbert term
\begin{equation}\label{LEH}
 L_{\rm EH}=-3M_{\rm Pl}^2\,a\,\dot a^2,
\end{equation}
which belongs to the first family of models\footnote{It must be noted the E--H term is not able to self-tune by itself \cite{Martin-Moruno:2015bda}.}
\begin{equation}
 L_{\rm linear}=a^3\sum_{i=0}^{3}\left[3\sqrt{\Lambda} \,U_i(\phi)+\dot\phi\,W_i(\phi)\right]\,H^i.
\end{equation}
 The second family of models has a Lagrangian given by 
\begin{equation}\label{Lnl}
L_{\rm nl}=a^3 \sum_{i=0}^{3}X_i\left(\phi,\,\dot\phi\right)H^i.
\end{equation}
The functions $U_i(\phi)$, $W_i(\phi)$, and $X_i\left(\phi,\,\dot\phi\right)$ are arbitrary up to the following constraints
which ensure the existence of a de Sitter critical point characterized by $\Lambda$, the late-time value of the Hubble rate,  for any kind of material content \cite{Martin-Moruno:2015bda}.
These are:
\begin{equation}\label{condition1}
  \sum_{i=0}^{3}W_i(\phi)\Lambda^{i/2}=\sum_{j=0}^{3}U_{j,\phi}(\phi)\Lambda^{j/2},
\end{equation}
and
\begin{equation}\label{condition2}
 \sum_{i=0}^{3} X_i\left(\phi,\,\dot\phi\right)\Lambda^{i/2}=0.
\end{equation}
The first constraint ensures that the Lagrangian density evaluated at the critical point has the form required to allow the field
to self-tune, which is
\begin{equation}\label{Lcp}
 \mathcal{L}_{\rm cp}=3\sqrt{\Lambda}\,h(\phi)+\dot\phi\, h_{,\phi}(\phi),
\end{equation}
with $h(\phi)$ an arbitrary function.
The second constraint forces any non-linear dependence of the Lagrangian on $\dot \phi$ to vanish at the critical point to agree with 
equation (\ref{Lcp}).
Furthermore, we consider that the material content is minimally coupled to the metric and non-interacting with the field. Thus, we have
\begin{equation}\label{Lm}
 L_{\rm m}= -a^3\rho(a)=-a^3\sum_s\rho_{s}(a),\qquad {\rm with}\qquad \rho_{s}(a)=\rho_{s,0}a^{-3(1+w_s)}.
\end{equation}

The cosmology of the linear models was considered in reference \cite{Martin-Moruno:2015lha}. In the present article we instead 
study the cosmology of the non-linear models, that is, we assume that 
$U_i(\phi)=W_i(\phi)=0$.
The field equation for the resulting models is given by \cite{Martin-Moruno:2015bda}
\begin{eqnarray}\label{field}
\sum_{i=0}^{3}   \left[X_{i,\phi}-3X_{i,\dot\phi}H- iX_{i,\dot\phi}\frac{\dot H}{H}-
 X_{i,\dot\phi \phi}\dot\phi-X_{i,\dot\phi\dot\phi}\ddot\phi\right] H^i=0.
\end{eqnarray}
where one needs at least one $i$ such that $X_{i,\dot\phi}\neq0$ with $i\neq0$ to obtain a dynamical self-tuning, that is, a term depending on $\dot H$ in the field equation. 
Moreover, taking into account condition (\ref{condition2}) and its derivatives, it can be checked that this equation trivially vanishes at the critical point, that is, when $H \rightarrow\sqrt{\Lambda}$.
On the other hand, the Hamiltonian density is
\begin{equation}\label{Htodo}
 \mathcal{H}=\mathcal{H}_{\rm EH}+\mathcal{H}_{\rm nl}+\mathcal{H}_{\rm m}=0,
\end{equation}
where
\begin{eqnarray}
\label{HEH}
 \mathcal{H}_{\rm EH}&=&-3M_{\rm Pl}^2H^2, \\
\label{Hnl}
\mathcal{H}_{\rm nl}&=&\sum_{i=0}^{3}\left[(i-1)X_i\left(\phi,\,\dot\phi\right)+\dot\phi \,X_{i,\dot\phi}\left(\phi,\,\dot\phi\right)\right]H^i\equiv\rho_\phi, \\
\label{Hm}
 \mathcal{H}_{\rm m}&=&\rho(a),
\end{eqnarray}
respectively, and we have defined $\rho_\phi$ for obvious reasons.
Since using (\ref{condition2}) and its derivatives gives
\begin{equation}
 \mathcal{H}_{\rm nl,\,os}=\sum_{i=0}^{3}i\, X_i\left(\phi,\,\dot\phi\right)\Lambda^{i/2},
\end{equation}
and we are assuming at least one $X_{i,\dot\phi}\neq0$ with $i\neq0$, 
this implies that 
\begin{equation}\label{conditionb}
 \sum_{i=0}^{3}i\, X_{i,\dot\phi}\Lambda^{i/2}\neq0,
\end{equation}
and consequently, $\mathcal{H}_{\rm nl,\dot\phi}\neq0$ at the critical point. Thus, the field is able to self-tune \cite{Martin-Moruno:2015bda}.
The modified Friedmann equation is given by (\ref{Htodo}). It can be explicitly written as
\begin{equation}\label{Friedmann}
 H^2=\frac{1}{3M_{\rm Pl}^2}\left(\rho+\rho_\phi\right).
\end{equation}
If we define $\Omega=\rho/(3M_{\rm Pl}^2H^2)$ and $\Omega_\phi=\rho_\phi/(3M_{\rm Pl}^2H^2)$, with $\rho_\phi$ given by equation (\ref{Hnl}), we can rewrite the modified Friedmann equation as
\begin{equation}\label{Friednl}
 \Omega+\Omega_\phi=1.
\end{equation}
Finally, one can define an equation of state parameter associated with the field as $w_\phi=p_\phi/\rho_\phi$, which can be written as 
\begin{equation}\label{wphi}
 w_\phi=-1-\frac{1}{3H}\frac{\dot\rho_\phi}{\rho_\phi},
\end{equation}
and an effective equation of state parameter as
\begin{equation}\label{weff}
 w_{\rm eff}=-1-\frac{2}{3}\frac{\dot H}{H^2}.
\end{equation}

\subsection{Shift-symmetric case}\label{shift}
Let us now focus our attention on models that are invariant under shift redefinitions of the field, $\phi\rightarrow\phi+c$.
In this case the functions $X_i$ are just a function of the field derivative, $X_i(\dot\phi)$. Let us  redefine the $X_i$ functions as
\begin{equation}\label{fs}
 f_i(\dot\phi)=\frac{\Lambda^{i/2-1}}{3M_{\rm Pl}^2}X_i(\dot\phi),
\end{equation}
for later convenience.
In terms of these new functions, conditions (\ref{condition2}) and (\ref{conditionb}) are respectively given by
\begin{equation}\label{condition}
 \sum_{i=0}^{3} f_i(\dot\phi)=0,\qquad {\rm and}\qquad \sum_{i=0}^{3}if_{i,\dot\phi}(\dot\phi)\neq0.
\end{equation}
As the Lagrangian is invariant under $\phi\rightarrow\phi+c$, there is a conserved quantity $\Sigma_0$ with
\begin{equation}\label{conserved}
 \sum_{i=0}^{3}f_{i,\dot\phi}(\dot\phi)h^i=\frac{1}{a^3}\frac{\partial L}{\partial \dot\phi}=\Sigma_0/a^3,
\end{equation}
where $h=H/\sqrt{\Lambda}$.
If $\Sigma_0=0$, taking into account condition (\ref{condition}), one can conclude that the universe stays either in a Minkowski state, $h=0$, or in a de Sitter state with, $h=1$ (as $\sum_{i=0}^{3}f_{i,\dot\phi}=0$), 
throughout the whole evolution.
For the more general case with $\Sigma_0\neq0$, one can obtain from equation (\ref{conserved}) that either $h\rightarrow0$ or $h\rightarrow1$ when $a\rightarrow\infty$.
Thus, for these models a late time de Sitter evolution is necessarily characterized by $h=1$ ($H=\sqrt{\Lambda}$).

As the functions $f_i(\dot\phi)$ only depend on $\dot\phi$, the field equation is greatly simplified and can be expressed in terms of $\psi\equiv\dot\phi$. Defining now, $N=\ln a$, as the new time variable, 
and denoting with a prime the derivatives with respect to $N$, this equation can be written as
\begin{equation}\label{fieldshift}
 3h \sum_{i=0}^{3}f_{i,\psi}(\psi)h^i+h\psi' \sum_{i=0}^{3}f_{i,\psi\psi}(\psi)h^i+h'\sum_{i=0}^{3}if_{i,\psi}(\psi)h^i=0.
\end{equation}
The modified Friedmann equation is just the constraint (\ref{Friednl}), which can be expressed as
\begin{equation}\label{Friedshift}
 \Omega+\Omega_\psi=1,
\end{equation}
where
\begin{equation}\label{Omf}
 \Omega_\psi=\sum_{i=0}^{3}\left[(i-1)f_i(\psi)+\psi f_{i,\psi}(\psi)\right]h^{i-2}.
\end{equation}
Finally, as each specie is conserved, we have
\begin{equation}\label{conserv}
 \Omega=\sum_s\Omega_{s},\qquad {\rm with}\qquad \Omega_{s}'=-\Omega_s\left[3(1+w_s)+2\frac{h'}{h}\right].
\end{equation}
Thus, the total density parameter satisfies the following differential equation 
\begin{equation}\label{conservtot}
 \Omega'=-\Omega\left[3(1+w)+2\frac{h'}{h}\right],
\end{equation}
where we have defined the total equation of state parameter of matter fluids as
\begin{equation}
 1+w=\frac{\sum_s\Omega_{s}(1+w_s)}{\sum_s\Omega_{s}},
\end{equation}
which should not be confused with $w_{\rm eff}$.
This quantity, $w$, generically is a function of the scale factor whenever there is more than one specie. For the cases when $w_s$ is a constant,  
equation (\ref{Omf}) can be integrated to yield
\begin{equation}\label{Os}
 \Omega_s=\frac{\Omega_{s,*}\,h_*^2}{h^2}\exp\left[-3(1+w_s)(N-N_*)\right],
\end{equation}
where the subscript $*$ means evaluation at the initial condition.

\section{Dynamical system analysis for shift-symmetric cases}\label{dynamical}

Considering the equations presented in the previous section, it can be noted that
we have three functions of $N$, those are $\psi$, $\Omega$, and $h$, which appear with their first derivatives in the field equation (\ref{fieldshift}), the Friedmann equation (\ref{Friedshift}),
and the conservation equation (\ref{conservtot}). Moreover, the Friedmann equation (\ref{Friedshift}) does not contain derivatives of any of these quantities, thus it can be seen as a constraint which
can be used to decouple the dependence on the derivatives of equation (\ref{fieldshift}) and (\ref{conservtot}).

With this aim let us first rewrite the field equation (\ref{fieldshift}) as
\begin{equation}\label{Eq1}
 \psi'P_1\left(h,\,\psi\right)+h'P_2\left(h,\,\psi\right)+P_0\left(h,\,\psi\right)=0,
\end{equation}
with
\begin{eqnarray}
P_0\left(h,\,\psi\right)&=&3h\sum_{i=0}^{3}f_{i,\psi}h^i\label{P0}\\
P_1\left(h,\,\psi\right)&=&h\sum_{i=0}^{3} f_{i,\psi\psi}h^i\label{P1}\\
P_2\left(h,\,\psi\right)&=&\sum_{i=0}^{3}if_{i,\psi}h^i.\label{P2}
\end{eqnarray}
Given that $\Omega=1-\Omega_\psi$ with $\Omega_\psi$ given by equation (\ref{Omf}), and substituting the resulting expression and its derivative in equation (\ref{conservtot})
we obtain
\begin{equation}\label{Eq2}
  \psi'Q_1\left(h,\,\psi\right)+h'Q_2\left(h,\,\psi\right)+3(1+w)Q_0\left(h,\,\psi\right)=0,
\end{equation}
where
\begin{eqnarray}
Q_0\left(h,\,\psi\right)&=&-1+\sum_{i=0}^{3}\left[(i-1)f_i+\psi f_{i,\psi}\right]h^{i-2}=\Omega_\psi-1\label{Q0}\\
Q_1\left(h,\,\psi\right)&=&\sum_{i=0}^{3}\left[if_{i,\psi}+\psi f_{i,\psi\psi}\right]h^{i-2}\label{Q1}\\
Q_2\left(h,\,\psi\right)&=&h^{-1}\left\{-2+\sum_{i=0}^{3}i\left[(i-1)f_i+\psi f_{i,\psi}\right]h^{i-2}\right\}.\label{Q2}
\end{eqnarray}
From equation (\ref{Eq1}) and (\ref{Eq2}) one can obtain
\begin{equation}\label{h'}
 h'=\frac{3(1+w)Q_0P_1-Q_1P_0}{Q_1P_2-Q_2P_1},
\end{equation}
\begin{equation}\label{psi'}
 \psi'=\frac{3(1+w)Q_0P_2-Q_2P_0}{Q_2P_1-Q_1P_2}.
\end{equation}
Assuming a single-component universe with $w$ constant, these two equations form an autonomous closed system. 
Taking into account (\ref{condition}), it can be noted that
$P_0\left(1,\,\psi\right)=P_1\left(1,\,\psi\right)=0$ and $P_2\left(1,\,\psi\right)=Q_1\left(1,\,\psi\right)\neq0$, and, 
therefore, $h'=0$ and $\psi'=-3(1+w)Q_0/Q_1$ at $h=1$.
Thus, we have a de Sitter critical point for any material content (that is, any $w$) provided $Q_0\left(1,\,\psi_c\right)=0$ and
$Q_1\left(1,\,\psi_c\right)\neq0$. The critical point is, therefore, characterized by
\begin{equation}\label{critical}
 \{h=1,\,\psi=\psi_c\},\qquad{\rm with}\qquad \sum_{i=1}^{3}i\,f_{i}\left(\psi_c\right)=1,
\end{equation}
obtained by demanding $Q_0(1,\psi_c) = 0$ and using the lhs relation of (\ref{condition}) and its first derivative, 
and
\begin{equation}
 \sum_{i=1}^{3}i\,f_{i,\psi}\left(\psi_c\right)\neq0,
\end{equation}
which results from requiring $Q_1\left(1,\,\psi_c\right)\neq0$ and using now the the second derivative of the lhs of (\ref{condition}).
Taking into account equations (\ref{Friedshift}) and (\ref{Q0}) it can be seen that this corresponds to $\Omega_c=0$, as it should be expected.

Let us now study the stability of the critical point using linear stability theory. The eigenvalues of the Jacobian matrix of the system given by equation (\ref{Eq1}) and (\ref{Eq2})
evaluated at the critical point (\ref{critical}) are
\begin{equation}
 \lambda_1=-3,\qquad\qquad\lambda_2=-3(1+w).
\end{equation}
Thus, the critical point is an attractor if $1+w>0$, a saddle point if $1+w<0$, and one should go beyond the linear stability analysis for pure vacuum. 
It must be emphasized that in this section we are considering a single component universe with constant $w$. 
Nevertheless, this result suggests the stability of more general models with components satisfying the null energy condition.

\section{Recovering early time GR with functions of the same order}\label{hlargo}

In order to understand which class of models can deliver an interesting cosmological behaviour,
we assume in this section, that {\it the functions $f_i(\psi)$ either vanish or are of the same order at early times}. 
In particular, we seek to identify which models possess a general relativistic cosmological history at times corresponding to very large values of the Hubble parameter, $h \gg1$.
One can note that,  due to constraint (\ref{condition}), models with only two non-vanishing functions, these are automatically of the same order as we must have $f_i(\psi)=-f_j(\psi)$.
Taking into account equations (\ref{weff}) and (\ref{h'}), the effective equation of state parameter is
\begin{equation}\label{1weff}
 1+w_{\rm eff}=-\frac{2}{3h}\frac{3(1+w)Q_0P_1-Q_1P_0}{Q_1P_2-Q_2P_1}.
\end{equation}
Thus, the referred assumption allow us to study the regime $h\gg1$ by keeping only the terms which are higher order in $h$
in the functions $Q_i$'s and $P_i$'s. We can classify the different cases by considering the non-vanishing $f_i$ with the largest value of $i$.

\paragraph{Case I.}
One can see that if the non-vanishing function with the largest value of $i$ is $f_3(\psi)$, 
one can approximate equation (\ref{1weff}) by
\begin{equation}
  1+w_{\rm eff}\simeq\frac{2}{3}\frac{(1+w)\left(2 f_3+\psi f_{3,\psi}\right)f_{3,\psi\psi}-\left(3f_{3,\psi}+\psi f_{3,\psi\psi}\right)f_{3,\psi}}{\left(2f_3+\psi f_{3,\psi}\right)f_{3,\psi\psi}-\left(3f_{3,\psi}+\psi f_{3,\psi\psi}\right)f_{3,\psi}},
\end{equation}
for $h\gg1$. Thus, assuming $0<w<1$ one has two different extremal cases, either
\begin{equation}
   1+w_{\rm eff}\simeq\frac{2}{3}(1+w),\qquad{\rm for}\qquad |\left(2 f_3+\psi f_{3,\psi}\right)f_{3,\psi\psi}|\gg|\left(3f_{3,\psi}+\psi f_{3,\psi\psi}\right)f_{3,\psi}|,
\end{equation}
or
\begin{equation}
   1+w_{\rm eff}\simeq\frac{2}{3},\qquad{\rm for}\qquad |\left(2 f_3+\psi f_{3,\psi}\right)f_{3,\psi\psi}|\ll|\left(3f_{3,\psi}+\psi f_{3,\psi\psi}\right)f_{3,\psi}|.
\end{equation}
Neither of these two cases is suitable to describe our Universe as radiation and later non-relativistic matter are expected to drive its dynamics at early times. 
One cannot obtain intermediate cases compatible with our cosmological history without any fine-tuning either.


\paragraph{Case II.}
Let us now consider that the non-vanishing function with the largest value of $i$ is $f_2(\psi)$.
Considering the regime $h\gg 1$ in equation (\ref{1weff}), we have
\begin{equation}
  1+w_{\rm eff}\simeq\frac{(1+w)\left(1-f_2-\psi f_{2,\psi}\right)f_{2,\psi\psi}+\left(2f_{2,\psi}+\psi f_{2,\psi\psi}\right)f_{2,\psi}}{\left(1-f_2-\psi f_{2,\psi}\right)f_{2,\psi\psi}+\left(2f_{2,\psi}+\psi f_{2,\psi\psi}\right)f_{2,\psi}}.
\end{equation}
For $0<w<1$ the extreme regimes are
\begin{equation}\label{condi21}
   1+w_{\rm eff}\simeq 1+w,\qquad{\rm for}\qquad |\left(1-f_2-\psi f_{2,\psi}\right)f_{2,\psi\psi}|\gg|\left(2f_{2,\psi}+\psi f_{2,\psi\psi}\right)f_{2,\psi}|,
\end{equation}
and
\begin{equation}
   1+w_{\rm eff}\simeq 1,\qquad{\rm for}\qquad |\left(1-f_2-\psi f_{2,\psi}\right)f_{2,\psi\psi}|\ll|\left(2f_{2,\psi}+\psi f_{2,\psi\psi}\right)f_{2,\psi}|.
\end{equation}
The case given by equation (\ref{condi21}) corresponds to a cosmological evolution which could be compatible with a general relativistic behaviour
at early times.
Analysing this case in further detail, we require a positive and small value of $\Omega_\psi$ for $h\gg1$. 
Taking into account equation (\ref{Omf}) one has that
\begin{equation}\label{condi22}
 \Omega_\psi\simeq f_2+\psi f_{2,\psi}\gtrsim0,
\end{equation}
for $h\sim h_*\gg1$ and $\psi\sim\psi_*$. 
The smallness of this quantity can be controlled by imposing a given field initial condition, $\psi_*$, 
when integrating equations (\ref{h'}) and (\ref{psi'}).
From inequality (\ref{condi22}) we must have $0<1-\left(f_2+\psi f_{2,\psi}\right)<1$ at early times, therefore, 
condition (\ref{condi21}) is satisfied if
\begin{equation}\label{condi23}
  |f_{2,\psi\psi}|\gg|\left(2f_{2,\psi}+\psi f_{2,\psi\psi}\right)f_{2,\psi}|,
\end{equation}
for $\psi\sim\psi_*$.
It is easier to ensure that conditions (\ref{condi22}) and (\ref{condi23}) are satisfied close to $\psi_*$ if 
the field only takes either positive or negative values.
A sufficient condition to guaranty this is to set a positive (negative) initial condition with $\psi'>0$ ($<0$).
Evaluating (\ref{psi'}) for $h\gg1$, and taking into account condition (\ref{condi21}), one has
\begin{equation}
 \psi'\simeq 3w\frac{f_{2,\psi}}{f_{2,\psi\psi}}.
\end{equation}
Thus, if $\psi_*>0$, one should require ${\rm sign}(f_{2,\psi})={\rm sign}(f_{2,\psi\psi})$, and ${\rm sign}(f_{2,\psi})\neq{\rm sign}(f_{2,\psi\psi})$ otherwise.
One has, of course, to check that if $\psi_*>0$ then $\psi_*<\psi_{\rm c}$, and $\psi_*>\psi_{\rm c}$ if $\psi_*<0$, with $\psi_{\rm c}$ given by equation (\ref{critical}), to ensure
that we are in the solution branch with the attractor.

\begin{figure}[!h]
\centering
\includegraphics[width=0.65\textwidth]{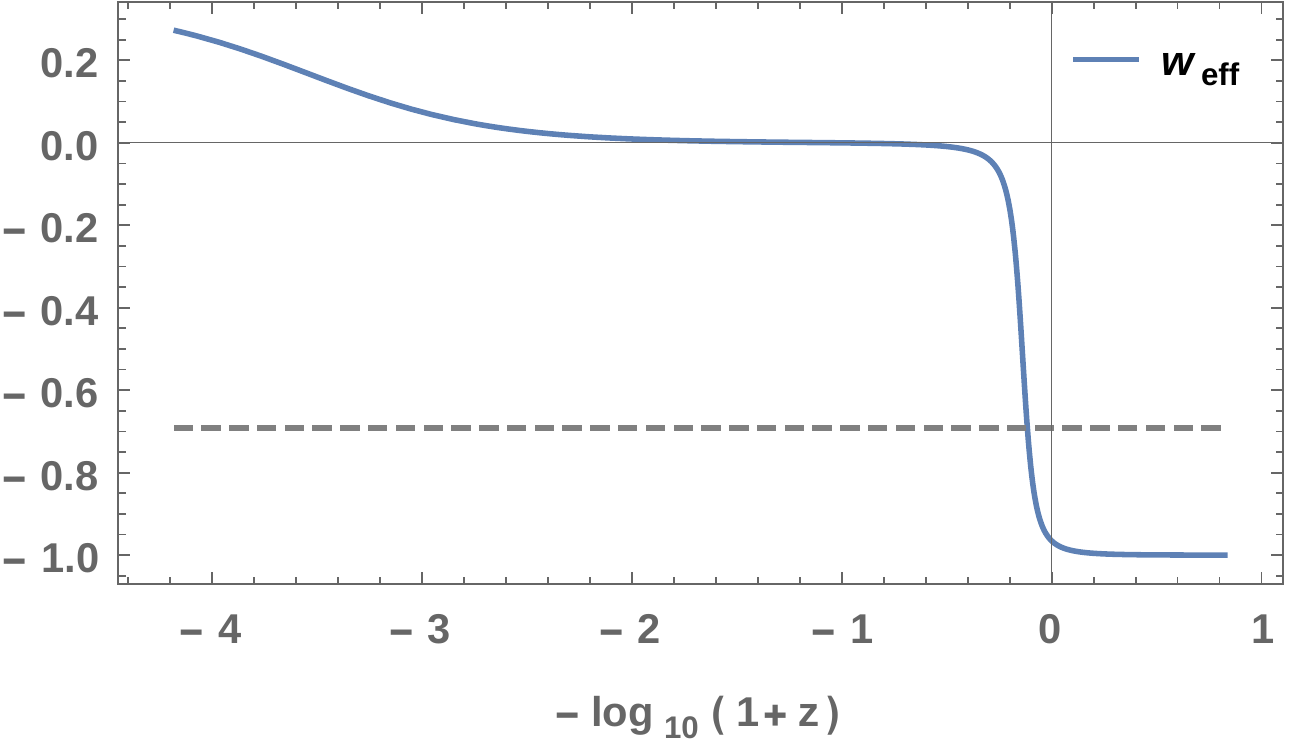}
\includegraphics[width=0.65\textwidth]{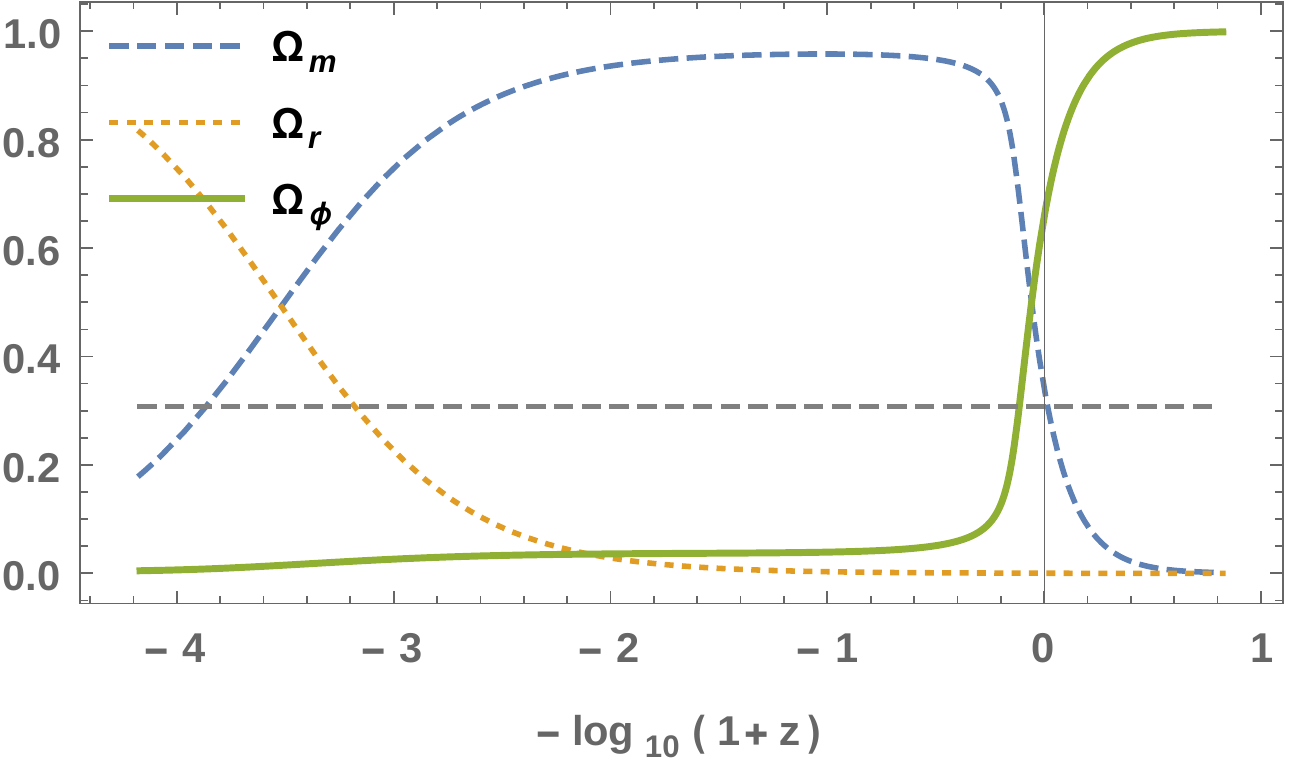}
\caption{Illustration  of a case II dynamics. Model given by the functions (\ref{exp}) with $a=10^{-2}$, $b=2$, $c=2a$ and $d=2b$. 
Initial conditions have been taken at equivalence, those are $z_*=z_{\rm eq}=3.3\times10^3$, $h_*=h_{\rm eq}=1.6\times10^5$, and $\psi_*=0.00001$.
On the upper panel we show the evolution of the effective equation
of state parameter for our model, depicting the current value favoured by observations. On the bottom  panel we show the evolution of 
the $\Omega$-parameters.}
\label{Fig1}
\end{figure}

\begin{figure}[!h]
\centering
\includegraphics[width=0.65\textwidth]{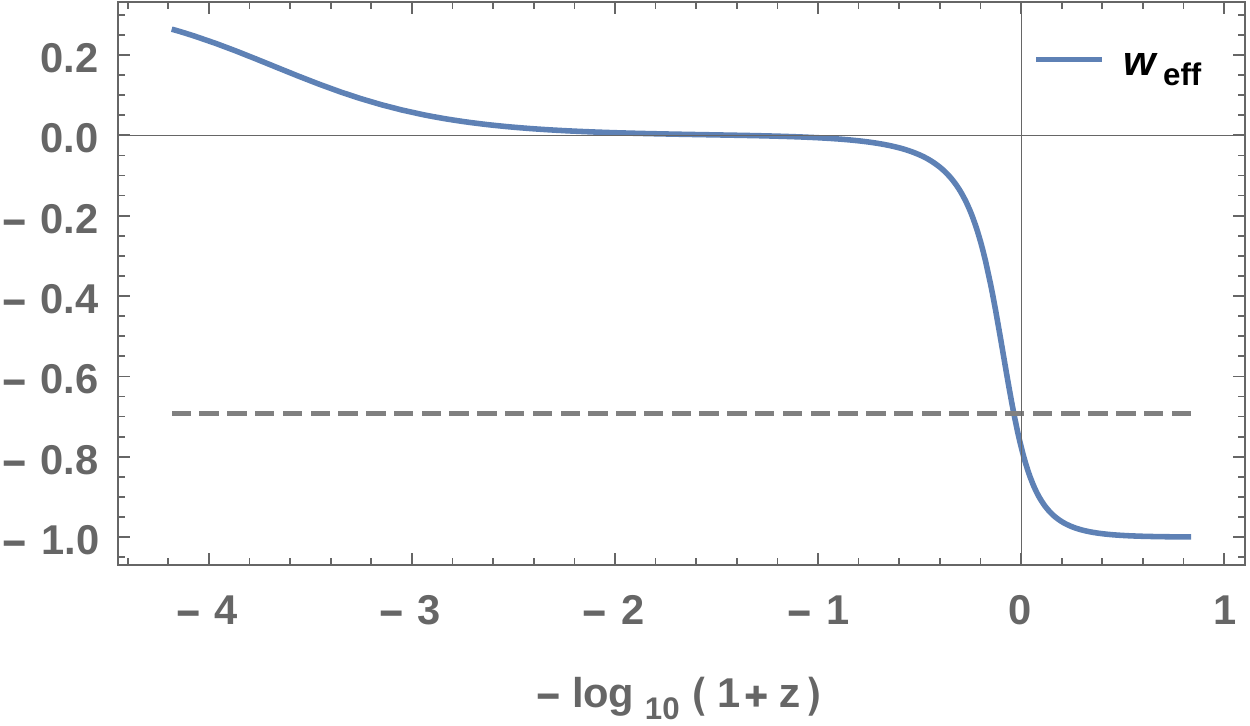}
\includegraphics[width=0.65\textwidth]{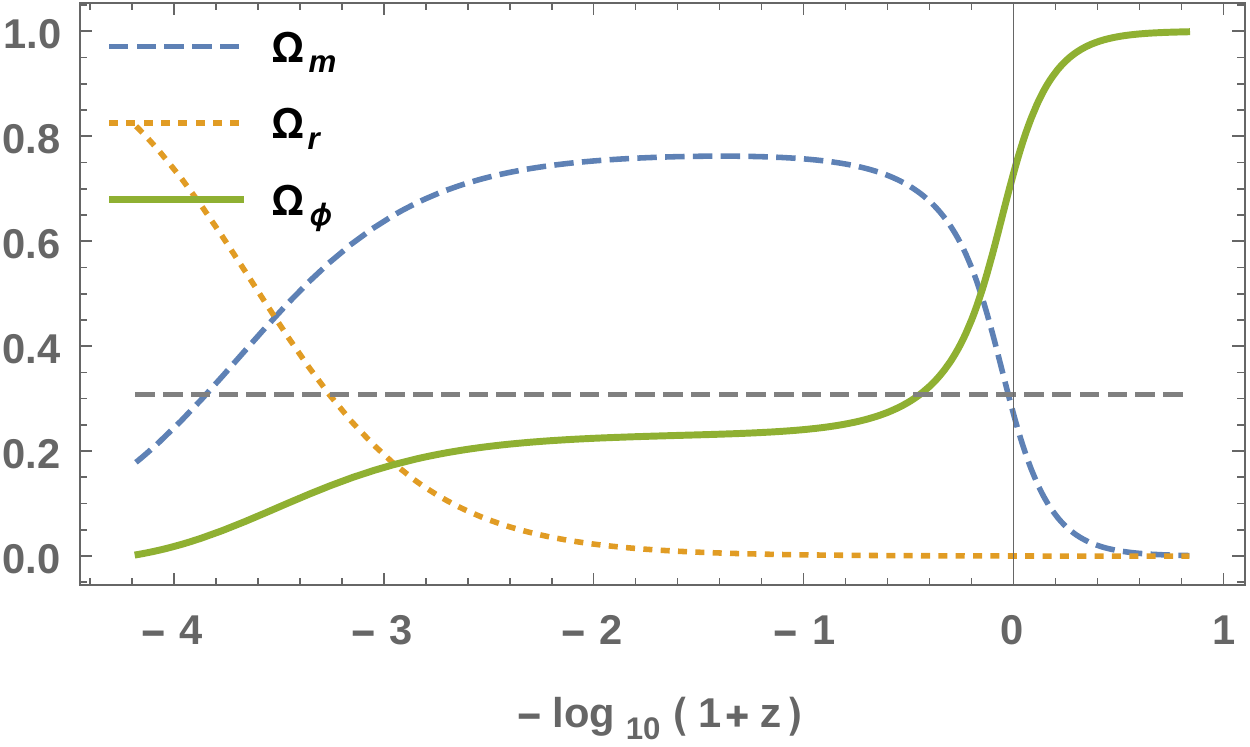}
\caption{Case II dynamics. Model given by the functions (\ref{exp}) with $a=9\times10^{-2}$, $b=2$, $c=2a$ and $d=2b$. 
Initial conditions have been taken at equivalence, those are $z_*=z_{\rm eq}=3.3\times10^3$, $h_*=h_{\rm eq}=1.6\times10^5$, and $\psi_*=0.00001$.}
\label{Fig2}
\end{figure}

These models are, therefore, very promising. Nevertheless, by studying different cases with power functions or exponential functions, we
have concluded that their late time behaviour is not completely satisfactory.
In figure \ref{Fig1} we show the result of the numerical integration of equations (\ref{h'}) and (\ref{psi'}) for a model described by 
\begin{equation}\label{exp}
f_2(\psi)=a\,e^{b\psi},\qquad f_1(\psi)=c\,e^{d\psi},\qquad f_0(\psi)=-a\,e^{b\psi}-c\,e^{d\psi},
\end{equation}
which has a behaviour similar to a model with power functions. 
It must be noted that the parameters of the model have to be such that conditions (\ref{condi22}) and (\ref{condi23}) are satisfied,
with $b$ and $d$ of the same order.
Both kind of models present a current value of $w_{\rm eff}$ too small to be compatible with observations,
as it can be seen in figure \ref{Fig1}.
One can only find a viable value for $w_{\rm eff,0}$ at the price of introducing large amounts of early dark energy. 
This is shown in figure \ref{Fig2}.


\paragraph{Case III.}
The remaining  case necessarily corresponds to the situation with only two potentials, $f_1(\psi)$ and $f_0(\psi)$. 
In the first place, taking into account equation (\ref{h'}) into equation (\ref{1weff}) for $h\gg1$, we get
\begin{equation}\label{caseiii}
 1+w_{\rm eff}\simeq1+w,
\end{equation}
for any form of the functions, making this models of particular interest.
In the second place, from equation (\ref{Omf}), we must impose
\begin{equation}\label{condi1}
 \Omega_\psi\simeq\frac{\psi f_{1,\psi}}{h}\gtrsim0,
\end{equation}
for $h\sim h_*\gg1$ and $\psi\sim\psi_*$, where we have implicitly assumed that $f_1/(\psi f_{1,\psi})$ evaluated at $\psi_*$ is not arbitrarily large.
The smallness of this quantity can be tuned by fixing the value of $\psi_*$ once a particular form of the function $f_i$ is given.
Condition (\ref{condi1}) is immediately satisfied if ${\rm sign}(\psi)={\rm sign}(f_{1,\psi})$, for $\psi\sim\psi_*$.
To ensure that $\psi$ does not change sign in the neighbourhood  of $\psi_*$ one can impose $\psi>0$ with $\psi'>0$ for $h\gg1$ or alternatively $\psi<0$ with $\psi'<0$. This is, of course, a sufficient condition although not necessary.
Considering the regime $h\gg1$ in equation (\ref{psi'}) one gets
\begin{equation}
 \psi'\simeq-\frac{3(1-w)f_{1,\psi}}{2f_{1,\psi\psi}}.
\end{equation}
Thus, for $0<w<1$, $\psi'>0$ ($<0$) if ${\rm sign}(f_{1,\psi})\neq{\rm sign}(f_{1,\psi\psi})$ (${\rm sign}(f_{1,\psi})={\rm sign}(f_{1,\psi\psi})$).
Taking into account also condition (\ref{condi1}) and the value of the field at the critical point (\ref{critical}), we can restrict our attention to consider models with either
$0<\psi_*<\psi_{\rm c}$, $f_{1,\psi}>0$ and $f_{1,\psi\psi}<0$, or $\psi_{\rm c}<\psi_*<0$, $f_{1,\psi}<0$ and $f_{1,\psi\psi}<0$, 
with $\psi_c$ given by equation (\ref{critical}).

\begin{figure}[!h]
\centering
\includegraphics[width=0.49\textwidth]{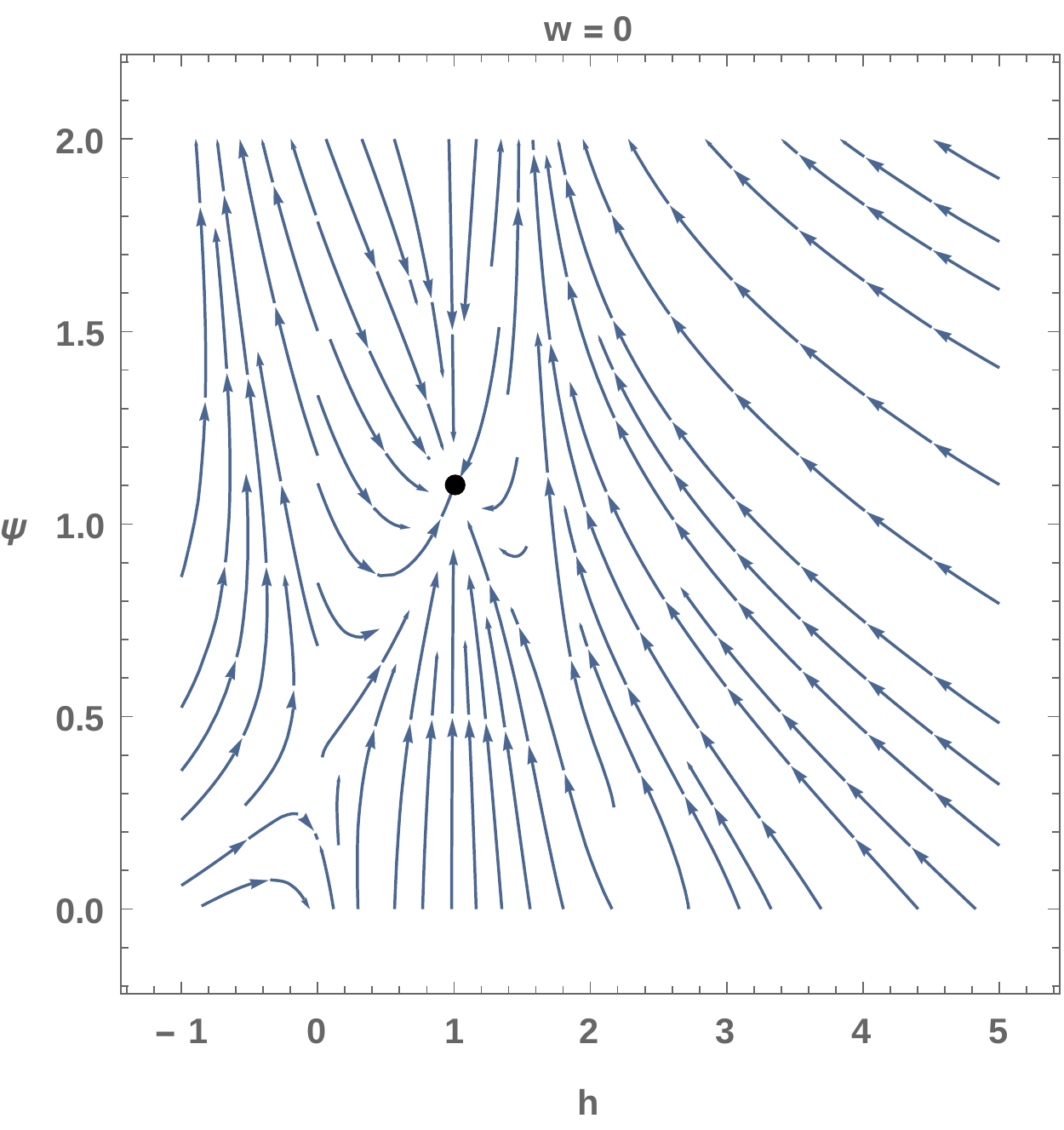}
\includegraphics[width=0.49\textwidth]{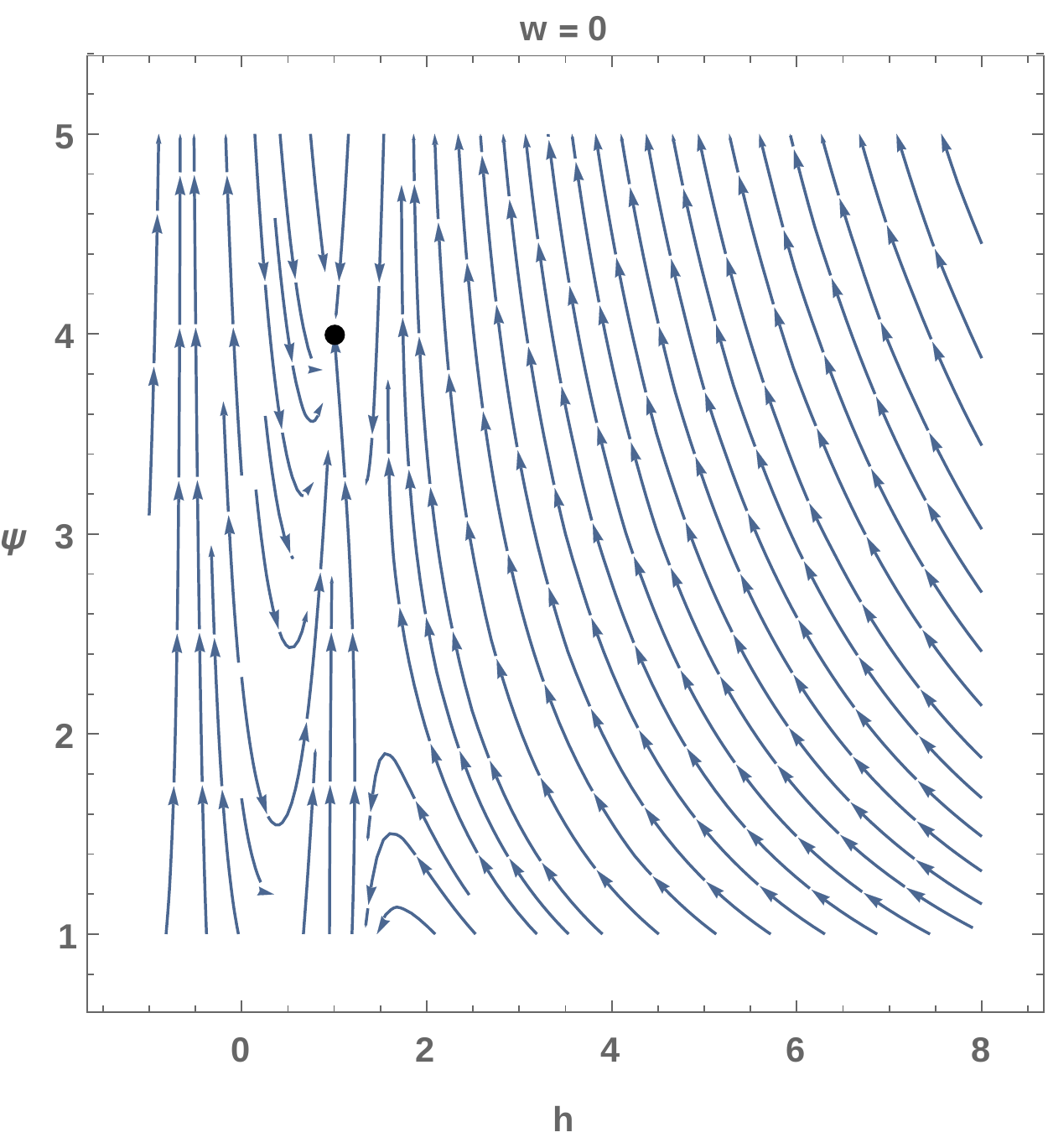}
\caption{Phase space diagram for case III. In the left panel we show the results for the model with a function given by equation (\ref{fi11}) with $c=2$, $b=3$ and $a=1$ and $d=2b$. 
In the right panel we depict the model corresponding to equation (\ref{fi12}) for $b=a=1/2$. 
The dynamics evolves away from the de Sitter critical point (indicated by the filled circle) in the branch of solutions
with large initial values of $h$.}
\label{Fig3}
\end{figure}

Despite the simplicity of these models, and encouraging characteristic shown in equation (\ref{caseiii}),
the arguments suggesting an early time general relativistic behaviour cannot be taken for granted if unexpected features appear in the phase space.
This is precisely what we find when studying particular models of this kind.
Assuming $\psi>0$ one can consider a model given by 
\begin{equation}\label{fi11}
 f_1(\psi)=c-b\,e^{-a\psi},
\end{equation}
and a second model with 
\begin{equation}\label{fi12}
 f_1(\psi)=b\,\psi^a,
\end{equation}
being $f_0=-f_1$, $c,\,b,\,a>0$. 
It must be noted that for the first case one needs $1<c<1+b$ to have $\psi_c>0$. 
The bounds come from equation (\ref{critical}), since $c=1+b\,e^{-a\psi_c}<1+b$ for $a\psi_c>0$ and $e^{a\psi_c}=b/(c-1)>0$ with $b>0$.
Looking at the phase diagram of these models in figure \ref{Fig3},
one can conclude that even if the critical point is an attractor, 
solutions with a large value for the initial condition for the Hubble parameter, $h_*$, 
correspond to a branch for which $\psi$ continues to grow. Thus, cosmologies with viable initial conditions will not reach the de Sitter attractor.

\section{Beyond the simplest assumption}\label{models}

Under the assumption considered in the previous section, that is, that at early times the functions $f_i(\psi)$ either vanish or are of the same order, we have been able to obtain promising results for models in which the function with larger value of $i$ is $f_2(\psi)$. 
These models have interesting early time cosmology, even though the current value of the cosmological parameters differ from values suggested by the current observational data.
In an attempt to obtain a more satisfactory cosmology at present time we will go beyond the assumption taken in the previous section still considering that $f_2(\psi)$ is the non-vanishing functions with the largest value of $i$ and that it satisfies the conditions presented in
the previous section.
It must be noted that,  because of condition (\ref{condition}), we need to have two additional functions to avoid them to be of the same order as $f_2(\psi)$. 
We consider a model with only $f_2(\psi)$ and $f_1(\psi)$, with an extra term that modifies $f_1(\psi)$ such that it differs substantially from $f_2(\psi)$ at least in some range of $\psi$. Thus, we study the following model:
\begin{equation}\label{beyond}
f_2(\psi)=\alpha\psi^n,\qquad f_1(\psi)=-\alpha\psi^n+\frac{\beta}{\psi^m},\qquad f_0(\psi)=-\frac{\beta}{\psi^m}.
\end{equation}
Looking to the past, if $h$ increases faster than $\psi$ decreases, the terms $\psi^n$ dominate and
this models has a consistent early cosmology for small enough values of $\alpha$, as in case II of the previous section.
If $\psi$ decreases faster than $h$ increases back in time, however, the models could develop the same characteristic in the phase space 
as shown in figure \ref{Fig3}. (We have studied that this can happen for large values of $\alpha$.)
The functions appearing in the general Lagrangian (without restriction to the minisuperspace) of this model are included 
in appendix \ref{Hfunctions} and \ref{Dfunctions}, depending whether one considers the Horndeski formulation \cite{Horndeski:1974wa}
or the Deffayet {\it et al.}~expression \cite{Deffayet:2011gz}.

\begin{figure}[!h]
\centering
\includegraphics[width=0.65\textwidth]{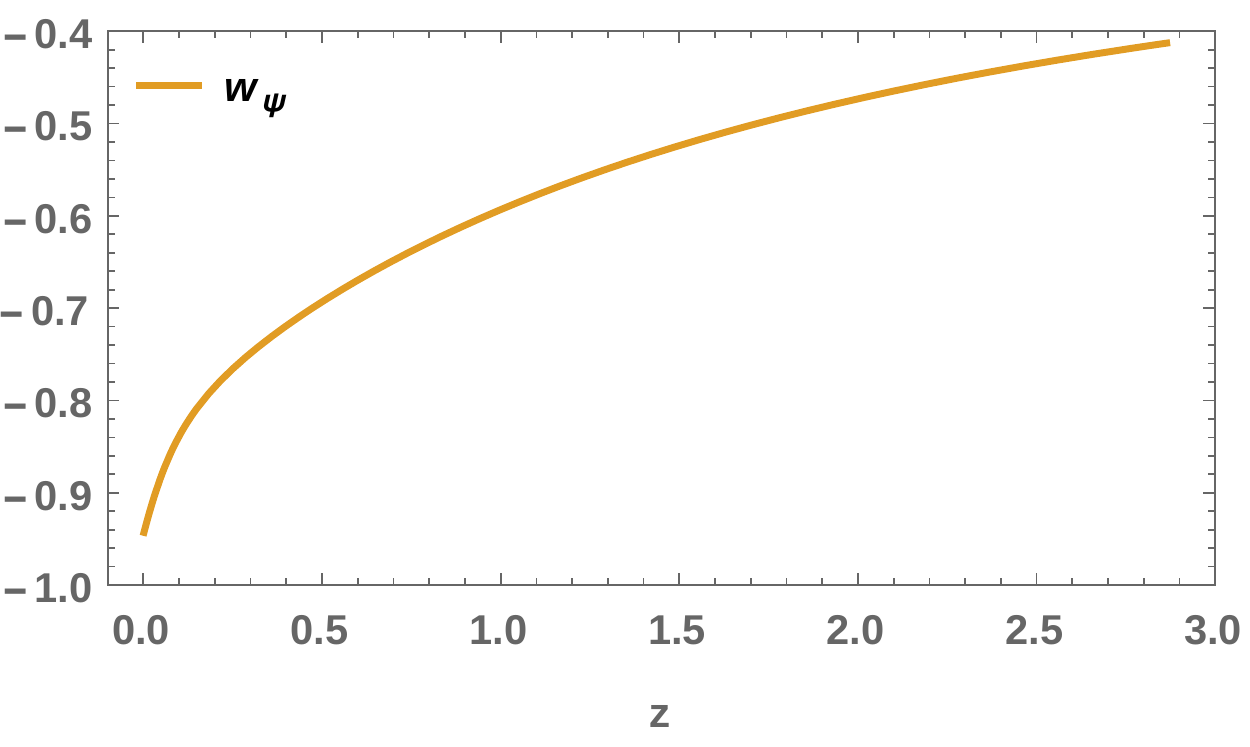}
\caption{Model given by the functions (\ref{beyond}) with parameters (\ref{casibueno}).
Although the evolution of other functions of the model is satisfactory, the decrease of the field equation of state parameter $w_\psi$ at
the present time, characterized by $w_a$, seems too large to easily agree with the observational data.}
\label{Fig4}
\end{figure}

\begin{figure}[!h]
\centering
\includegraphics[width=0.65\textwidth]{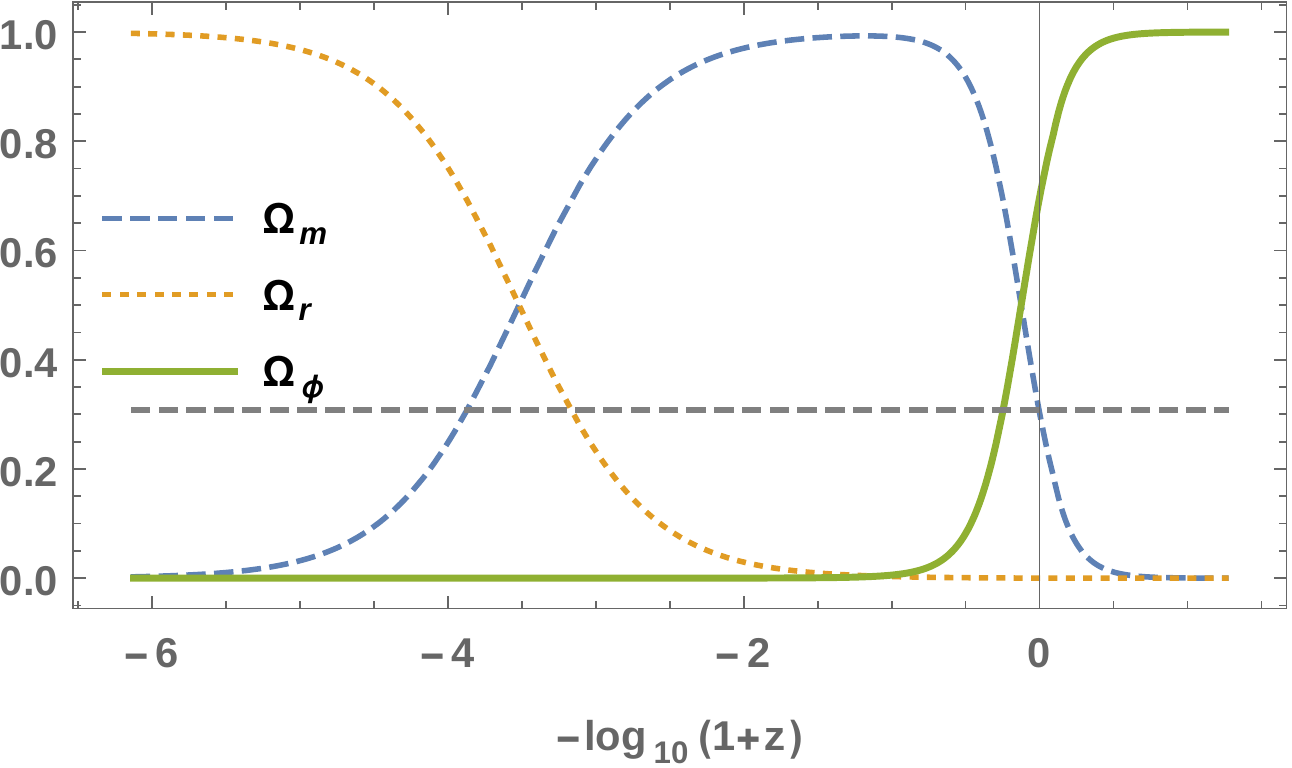}
\includegraphics[width=0.65\textwidth]{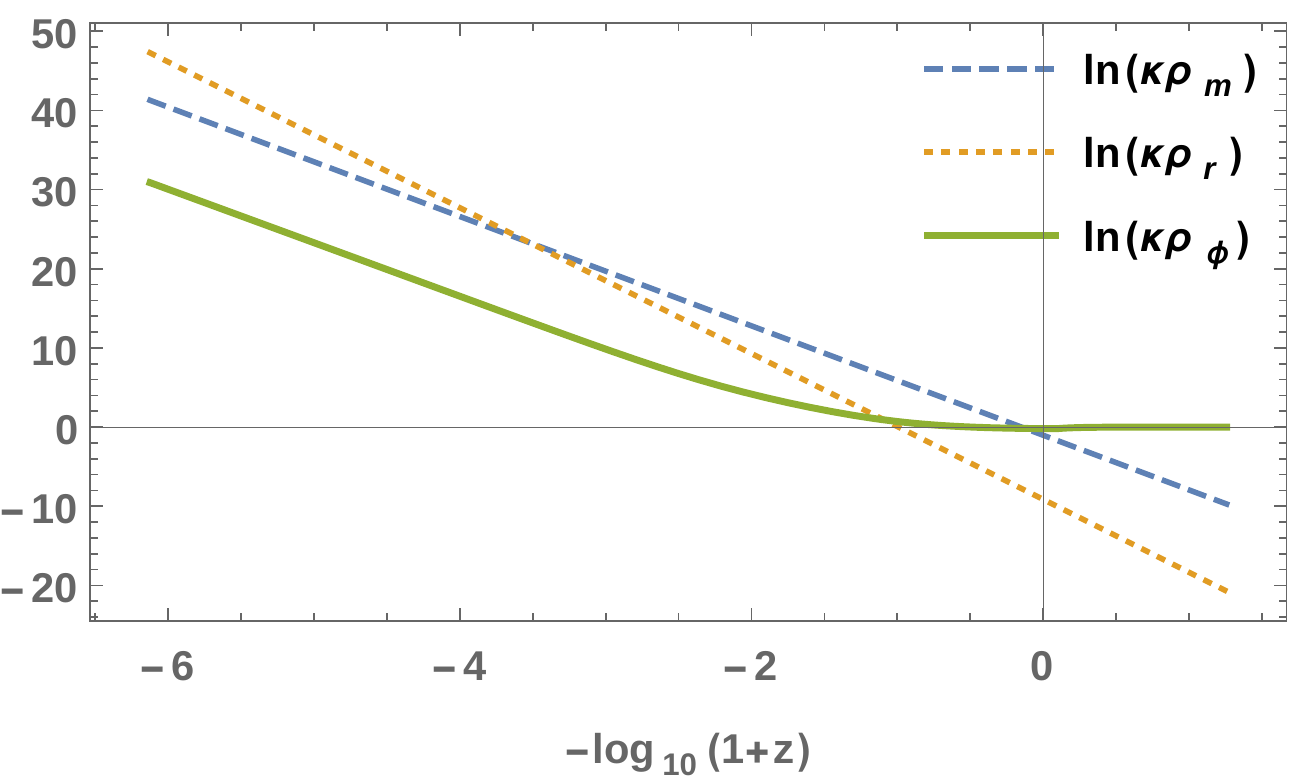}
\caption{Model given by the functions (\ref{beyond}) with parameters (\ref{bueno}).
On the top panel we compare the evolution of the $\Omega$-parameters, and on the bottom panel, 
the matter and field content in terms of the logarithmic of the energy densities, where $\kappa=8\pi\,G/(3\Lambda)$.}
\label{Fig5}
\end{figure}

\begin{figure}[!h]
\centering
\includegraphics[width=0.65\textwidth]{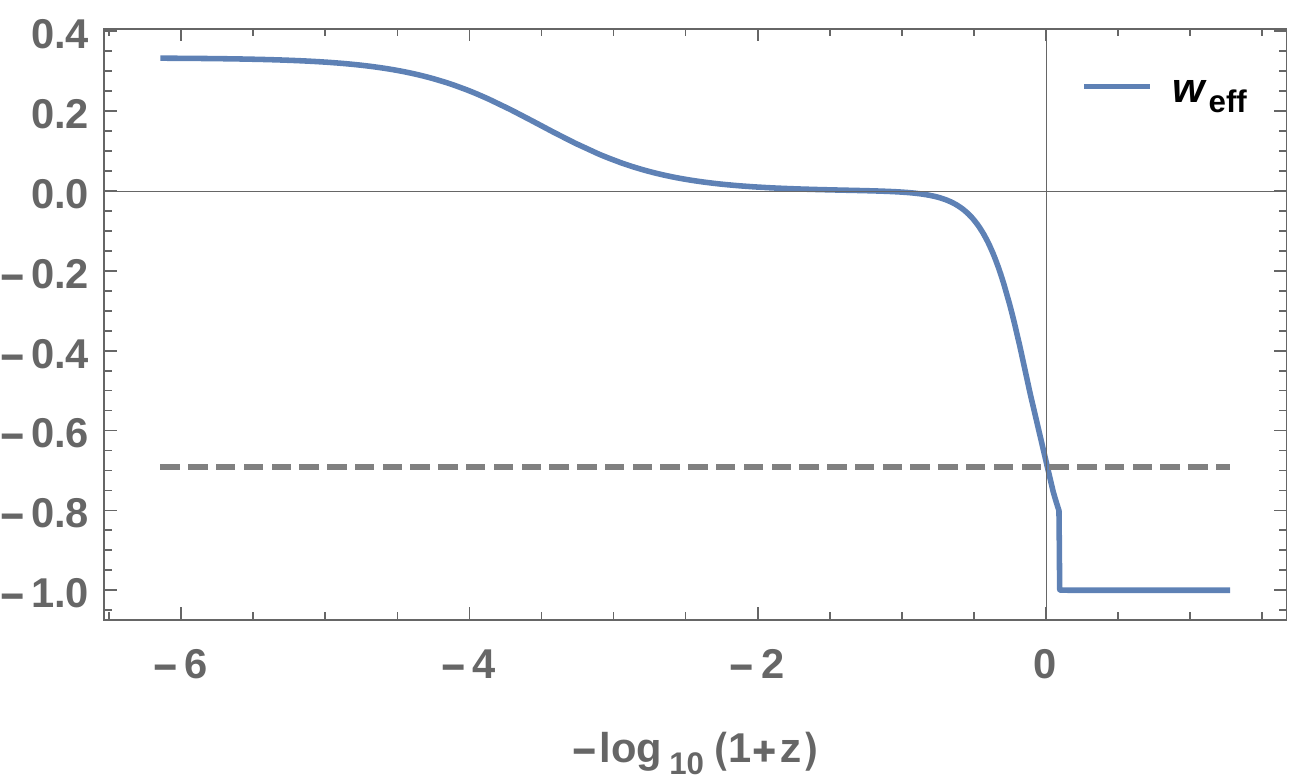}
\includegraphics[width=0.65\textwidth]{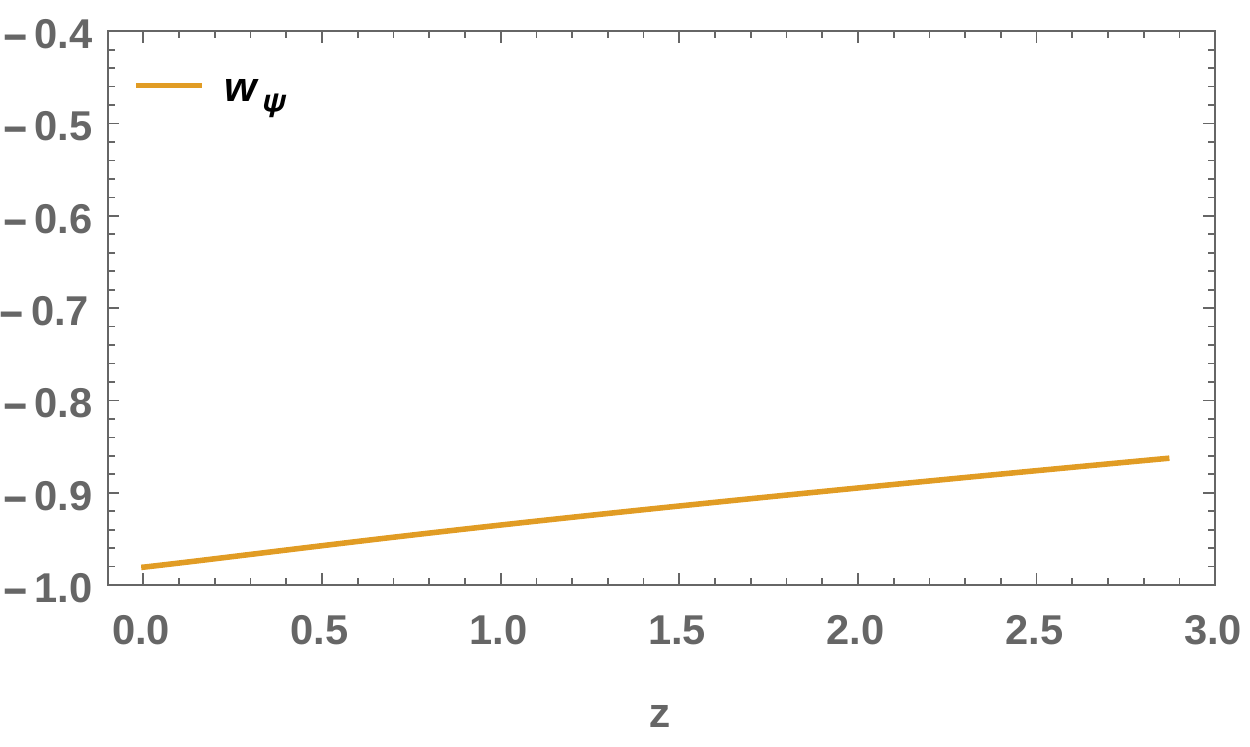}
\caption{Model given by the functions (\ref{beyond}) with parameters (\ref{bueno}).
On the top panel we show the evolution of the effective equation
of state parameter and the current value favoured by observations. On the bottom panel we show the evolution 
of the field equation of state parameter.}
\label{Fig6}
\end{figure}

Considering a universe filled with radiation and non-relativistic matter, one can numerically integrate equations (\ref{h'}) and (\ref{psi'}). 
Taking into account the evolution of the conserved matter content, equation (\ref{Os}), one can depict the relevant quantities of the model. 
The evolution of these models is quite satisfactory. 
For example, for a model with
\begin{equation}\label{casibueno}
 \alpha=10^{-3},\qquad n=2,\qquad m=1,\qquad \beta=7.5,
\end{equation}
taking $h_{\rm eq}=1.6\times10^{5}$ and $\psi_{\rm eq}=1.0010074751$ at equivalence 
$z_{\rm eq}=3.3\times10^3$,
we obtain consistent evolutions of the $\Omega$-parameters and current values compatible with observations,
$\Omega_{{\rm m,}0}=0.300929$ and $\Omega_{{\rm r,}0}=9.11907\times10^{-5}$, and small quantities of early dark energy, 
$\Omega_{\phi,{\rm rec}}=6.77705\times10^{-3}$. 
Nevertheless, considering a Taylor expansion of the field equation of state parameter, depicted in
figure \ref{Fig4},
around the current scale factor as
\begin{equation}\label{wa}
 w_\psi\simeq w_{0}+w_{a}(1-a),
\end{equation}
we obtain $w_0=-0.945417$ and $w_a=1.49429$, which are at the very best only in marginal agreement with the observational constraints (see e.g. \cite{Ade:2015rim}).
On the other hand, for a model with
\begin{equation}\label{bueno}
 \alpha=10^{-6},\qquad n=16,\qquad m=1,\qquad \beta=1.5,
\end{equation}
the cosmological history becomes entirely compatible with current data, considering conditions $h_{\rm eq}=1.6\times10^{5}$ and 
$\psi_{\rm eq}=1.03999$ at equivalence $z_{\rm eq}=3.3\times10^3$. The evolution of the energy densities for this particular case is shown 
in figure \ref{Fig5}. The results are compatible with current observational data, as we obtain $\Omega_{{\rm m,}0}=0.302601$
and $\Omega_{{\rm r,}0}=9.16972\times10^{-5}$, avoiding early dark energy with $\Omega_{\phi,{\rm rec}}=3.90298\times10^{-5}$.
In figure \ref{Fig6} we show the evolution of the effective equation of state parameter and the field equation of state parameter.
As we show, the current value of $w_{\rm eff}$ is compatible with the data. 
Moreover, we obtain for our model $w_0=-0.980623$ and $w_a=0.0436749$, equation (\ref{wa}). 
It can be verified that this model presents a future brief phantom epoch
before approaching the cosmological constant behaviour.

\section{Summary and further comments}\label{discussion}
In this article we have consider Horndeski cosmological models that may alleviate the cosmological constant problem by screening
any value of the vacuum energy given by the theory of particle physics. 
In particular, we have studied in detail the non-linear family of models obtained in reference \cite{Martin-Moruno:2015bda}
which have a de Sitter critical point for any material content. 
Furthermore, we have considered models protected by a particular symmetry, a shift symmetry of the field.

As we have shown, for these shift-symmetric models the de Sitter critical point is indeed an attractor. 
Thus, we can understand the current accelerated expansion of our Universe as the result 
of the dynamical approach of the field to  the critical point,
being the value of $H^2 \rightarrow \Lambda$ at this point completely independent of the vacuum energy.

The background cosmological evolution of the models studied in this article suggests that these models are in even better footing than the linear family considered in reference \cite{Martin-Moruno:2015lha}. 
It is clear from our analysis that there is a region of parameter space that it is incompatible with current observational data, hence, these models are susceptible of being ruled out. However, we have also
identified a particular case able to describe currently available observational data which depends of four parameters. 
Imposing the integration conditions at matter-radiation equality, this model provides us with a value
of the density parameters and equation of state field parameter at the present time 
within observational bounds. Moreover, the contribution of the field at early times
is negligible, recovering a cosmological dynamics compatible with that produced by general relativity.
In particular, there is no early dark energy.

In order to scrutinise these models we are now required to face them against observables that depend on the evolution of the field and matter fluid fluctuations. 
Another possible extension of the current work consists in investigating how the linear and non-linear contributions to the minisuperspace Lagrangian affect the dynamics when they are both present. 
This will be carried out in future work.

As we stated in the introduction, these models can  alleviate the cosmological constant problem only
 if a long enough radiation and matter phases can be described before the attractor is approached
 for any value of the vacuum energy.
 This point has not been addressed in the present work.
 One could expect the dynamical screening to start before complete screening has taken 
 place at the critical point.
 Moreover, for some models it may even be possible that the screening could be more effective 
 for vacuum energy than for the material content, allowing a consistent cosmology.
 Such study should, however, be carefully carried out in a follow up project.

\begin{acknowledgments}
The authors acknowledge Miguel Zumalacarregui for useful comments.
This work was supported by the Funda\c{c}\~{a}o para a Ci\^{e}ncia e Tecnologia (FCT) through 
the grants EXPL/FIS-AST/1608/2013 and UID/FIS/04434/2013. 
PMM also acknowledges financial support from the Spanish Ministry of Economy and Competitiveness 
through the postdoctoral training contract FPDI-2013-16161 and the project FIS2014-52837-P.
\end{acknowledgments}


\appendix

\section{Horndeski functions}\label{Hfunctions}
In this paper, we have started by including the Lagrangian already restricted to the minisuperspace, which is the case of interest for
studying the background cosmology of a particular model.
Nevertheless, once a particular satisfactory model has been found, one needs to write the general Lagrangian to study other consequences
of the model. In order to facilitate that study for future works, we include here the general Lagrangian of the model presented in section
\ref{models}.

The Horndeski Lagrangian, as expressed in reference \cite{Horndeski:1974wa}, can be written as
\begin{eqnarray}\label{H}
 \mathcal{L}_H&=&\delta^{\alpha\beta\gamma}_{\mu\nu\sigma}\left[\kappa_1\nabla^\mu\nabla_\alpha\phi \,R_{\beta\gamma}{}^{\nu\sigma}
 -\frac{4}{3}\kappa_{1,X}\nabla^\mu\nabla_\alpha\phi\nabla^\nu\nabla_\beta\phi\nabla^\sigma\nabla_\gamma\phi\right.\nonumber\\
 &+&\left.\kappa_3\nabla_\alpha\phi\nabla^\mu\phi\,R_{\beta\gamma}{}^{\nu\sigma}-
 4\kappa_{3,X}\nabla_\alpha\phi\nabla^\mu\phi\nabla^\nu\nabla_\beta\phi\nabla^\sigma\nabla_\gamma\phi\right]\nonumber\\
 &+&\delta_{\mu\nu}^{\alpha\beta}\left[F\,R_{\alpha\beta}{}^{\mu\nu}-4F_{,X}\nabla^\mu\nabla_\alpha\phi \nabla^\nu\nabla_\beta\phi
 +2\kappa_8\nabla_\alpha\phi\nabla^\mu\phi\nabla^\nu\nabla_\beta\phi\right]\nonumber\\
 &-&3\left[2F_{,\phi}+X\,\kappa_8\right]\nabla_\mu\nabla^\mu\phi+\kappa_9,
\end{eqnarray}
where 
\begin{equation}\label{condF}
 F_{,X}=\kappa_{1,\phi}-\kappa_3-2X\kappa_{3,X},
\end{equation}
$ X=\nabla_\mu\phi\nabla^\mu\phi$, and $\kappa_i\left(\phi,\,X\right)$ are arbitrary functions.
We are considering shift-symmetric models, therefore, the functions are just dependent on the kinetic term $\kappa_i\left(X\right)$. 
As it was first shown in reference \cite{Charmousis:2011ea} for the minisuperspace Lagrangian in the general case,
the functions of Lagrangian (\ref{Lnl}) are related with the functions appearing in Lagrangian (\ref{H}) through
\begin{eqnarray}
 X_0&=&-\widebar Q_{7,\phi}\dot\phi+\kappa_9,\label{X0}\\ 
 X_1&=&-3\,\widebar Q_7+\widebar Q_{7,\dot\phi}\dot\phi,\label{X1}\\
 X_2&=&12\,F_{,X}X-12\,F,\label{X2}\\ 
 X_3&=&8\,\kappa_{1,X}\,\dot\phi^3, \label{X3}
\end{eqnarray}
with
\begin{eqnarray}
\widebar Q_{7,\dot\phi}=-3\,\dot\phi^2\kappa_8,\label{Q7}
\end{eqnarray}
where we have simplified the expressions due to the shift-symmetry.
Thus, the Lagrangian for the models given by (\ref{beyond}) have the following Horndeski functions
\begin{eqnarray}
 \kappa_1&=&c_1, \label{kappa1}\\
 \kappa_3&=&\frac{M_{\rm Pl}^2\,n}{4(n-2)(n-1)} \alpha\left(-X\right)^{n/2-1}+c_2+\frac{c_3}{\left(-X\right)^{1/2}},   \\
 F&=&\frac{M_{\rm Pl}^2}{2(n-2)}\alpha\left(-X\right)^{n/2}-c_2X, \\
 \kappa_8&=&M_{\rm Pl}^2\sqrt{\Lambda}\left[\frac{n}{n-3}\alpha\left(-X\right)^{n/2-3/2}
 -\frac{m}{m+3}\beta\left(-X\right)^{-m/2-3/2}\right]-c_4,  \\
 \kappa_9&=&-3M_{\rm Pl}^2\,\Lambda \beta\left(-X\right)^{-m/2}\label{kappa9},
\end{eqnarray}
being the $c_i$'s integration constants. 
It must be noted that the terms appearing multiplied by $c_1$, $c_2$ and $c_4$ can be combined in
a total derivative; therefore,
these constants can be fixed to zero.
The constant $c_3$ does not appear in the minisuperspace Lagrangian, therefore, the corresponding term
is not able to self-tune to de Sitter by itself although it does not spoilt screening (as it happens
in the case of the linear models with two potentials \cite{Martin-Moruno:2015bda}).


\section{Deffayet et al. functions}\label{Dfunctions}

Deffayet {\it et al.} independently found the Horndeski Lagrangian in reference \cite{Deffayet:2011gz},
expressed in a form which is currently more used in the literature. Assuming a shift-symmetric field, this is
\begin{eqnarray}\label{LD}
 \mathcal{L}&=&K(X)-G_3(X)\,\square\phi+G_4(X)\,R-2\,G_{4,X}\left[(\square\phi)^2-(\nabla_\mu\nabla_\nu\phi)^2\right]\nonumber\\
 &+&G_5(X)G_{\mu\nu}\nabla^\mu\nabla^\nu\phi+\frac{1}{3}G_{5,X}\left[(\square\phi)^3-3(\square\phi)(\nabla_\mu\nabla_\nu\phi)^2+
 2(\nabla_\mu\nabla_\nu\phi)^3\right].
\end{eqnarray}
Taking  this  symmetry  into  account,  the  dictionary  first  presented  in  reference \cite{Kobayashi}
relating the former Lagrangian with Lagrangian (\ref{H}) can be expressed as
\begin{eqnarray}
 K&=&\kappa_9,\label{K}\\
 G_3&=&X\kappa_8-\int^X{\rm d}X'\kappa_{8},\label{G3}\\
 G_4&=&2F+2X\kappa_3,\label{G2}\\
 G_5&=&-4\kappa_1.\label{G1}
\end{eqnarray}
For the model given by equation (\ref{beyond}), taking into account equations (\ref{kappa1}-\ref{kappa9}), we have
\begin{eqnarray}
 K&=&-3M_{\rm Pl}^2\,\Lambda \beta\left(-X\right)^{-m/2}, \\
 G_3&=&M_{\rm Pl}^2\sqrt{\Lambda}\left[-\frac{n}{n-1}\alpha\left(-X\right)^{n/2-1/2}+\frac{m}{m+1}\beta\left(-X\right)^{-m/2-1/2}\right],\label{G3}\\
 G_4&=&\frac{M_{\rm Pl}^2}{2(n-1)}\alpha\left(-X\right)^{n/2}-2\,c\left(-X\right)^{1/2},\label{G2}\\
 G_5&=&0,  \label{G1}
\end{eqnarray}
where we do not take into account the terms leading to total derivatives, and we have defined
$c=2c_3$.
As $c$ does not affect the background cosmology, the value of this constant is not restricted by our 
analysis, and it can be fixed to the more convenient value. 
If one considers the limit case $\alpha=0$, one is in case III of section (\ref{hlargo}) for $m<0$.
This case is particularly simple as the Lagrangian (\ref{LD}) only contains two terms. On the other hand, the limit case $\beta=0$
corresponds to case II of section (\ref{hlargo}).


\end{document}